\documentclass[prl,amsfonts,amsmath,showpacs,twocolumn,floatfix]{revtex4}

\usepackage{bm}
\usepackage{graphicx}

\def\be{\begin{equation}}
\def\ee{\end{equation}}

\def\br{{\bf r}}

\newcommand{\corr}[1]{\langle #1\rangle}

\def\Re{\mathop{\rm Re}}
\def\Im{\mathop{\rm Im}}

\def\tr{\mathop{\rm tr}}

\def\str{\mathop{\rm str}}
\def\diag{\mathop{\rm diag}}

\def\erf{\mathop{\rm erf}}

\def\const{\text{const}}
\def\ETh{E_{\text{Th}}}

\def\mls{\delta}

\def\ucmatrix{\gamma}
\def\cicmatrix{\alpha}

\begin{document}

\title{Crossovers between superconducting symmetry classes}

\author{V. A. Koziy}
\affiliation{L. D. Landau Institute for Theoretical Physics, 142432 Chernogolovka, Russia}
\affiliation{Moscow Institute of Physics and Technology, 141700 Moscow, Russia}

\author{M. A. Skvortsov}
\affiliation{L. D. Landau Institute for Theoretical Physics, 142432 Chernogolovka, Russia}
\affiliation{Moscow Institute of Physics and Technology, 141700 Moscow, Russia}

\date{\today}

\begin{abstract}
We study the average density of states in a small metallic grain
coupled to two superconductors with the phase difference $\pi$,
in a magnetic field. The spectrum of the low-energy excitations
in the grain is described by the random matrix theory whose symmetry
depends on the magnetic field strength and coupling to the
superconductors.
In the limiting cases,
a pure superconducting symmetry class is realized.
For intermediate magnetic fields or couplings
to the superconductors, the system experiences a crossover
between different symmetry classes. With the help of the supersymmetric
$\sigma$-model we derive the exact expressions for the average density
of states in the crossovers between the symmetry classes A--C and CI--C.
\end{abstract}


\maketitle

\textbf{Introduction.}
Energy levels in small metallic particles with chaotic
electron dynamics are random numbers.
It is generally accepted that their spectral statistics
in the ergodic regime
is described by the random matrix theory (RMT) \cite{Mehta}.
For disordered grains, this had been proved by Efetov \cite{Efetov1983}
with the help of the supersymmetry technique \cite{Efetov-book},
while for quantum billiards in the absence of disorder
this statement is usually referred to
as the Bohigas conjecture \cite{Bohigas}.

In the RMT,
a system is characterized solely by its symmetry.
In the application to condensed matter, the standard three
Wigner-Dyson ensembles (orthogonal, unitary and symplectic)
\cite{Dyson-classes} describe level statistics in small metallic
grains in the presence or absence of the time-reversal and
spin-rotation symmetries \cite{Efetov-book}.

Recently, the Wigner-Dyson classification had been extended
to superconducting \cite{AZ97} and chiral \cite{chiral} symmetry classes,
which arise when the Hamiltonian possesses an additional symmetry
with respect to changing the sign of the energy
(counted from the Fermi energy).
With the appearance of a selected energy point,
in the superconducting/chiral classes even the average density
of states (DOS), $\corr{\rho(E)}$, becomes a nontrivial
function of the energy. This should be contrasted to the standard
Wigner-Dyson ensembles where
$\corr{\rho(E)} = \delta^{-1} = \const$ and
the first nontrivial quantity is the pair correlation function
$R_2(\omega) = \delta^2\corr{\rho(E+\omega)\rho(E)}-1$.

The symmetry classes (three Wigner-Dyson, four superconducting
and three chiral) correspond to the limits when various
symmetries are either present or completely broken.
In the intermediate cases,
the system experiences a crossover between different symmetry classes.
The pair correlation function in the crossover between
the orthogonal and unitary classes was obtained in Refs.~\cite{Pandey-Mehta-O-U,AIE93}:
\begin{multline}
\label{pair-corr-O-U}
  R_2^\text{O-U}(\omega)
  =
  1 - \frac{\sin^2x}{x^2}
\\{}
  +
  \int_1^\infty d\lambda \,
    \frac{\sin\lambda x}{\lambda} \, e^{-\alpha\lambda^2}
  \int_0^1 d\mu \, \mu \sin\mu x \, e^{\alpha\mu^2}
  ,
\end{multline}
where $x=\pi\omega/\delta$ and $\alpha$ is the symmetry-breaking
parameter.
Equation (\ref{pair-corr-O-U}) interpolates between
the orthogonal ($\alpha=0$) and unitary ($\alpha=\infty$) results.
The pair correlator in known also in the symplectic--unitary
crossover \cite{Mehta-Pandey-Sp-U,Kravtsov-Zirnbauer},
its form being similar to Eq.~(\ref{pair-corr-O-U}).

\begin{figure}
\centerline{
\includegraphics{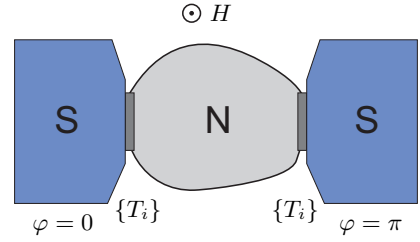}
}
\caption{A normal-metal dot coupled to two superconducting
terminals with the phase difference $\pi$, in a magnetic field.
NS interfaces are characterized by the set of transparencies $\{T_i\}$.}
\label{F:setup}
\end{figure}

\begin{figure}[b]
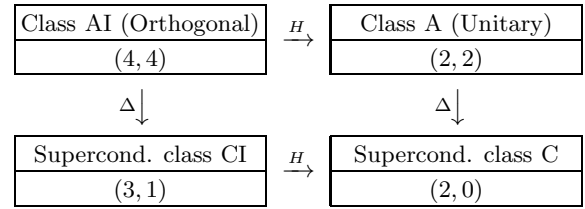

$$
\begin{array}{ccc}
\begin{tabular}{|c|}
\hline
\makebox[32mm]{Class AI (Orthogonal)}
\\
\hline
$(4,4)$ \\
\hline
\end{tabular}
&
\xrightarrow{H}
&
\begin{tabular}{|c|}
\hline
\makebox[32mm]{Class A (Unitary)}
\\
\hline
$(2,2)$ \\
\hline
\end{tabular}
\\[12pt]
\makebox[0pt][r]{\scriptsize $\Delta$}\rotatebox[origin=c]{-90}{$\longrightarrow$}
&
&
\makebox[0pt][r]{\scriptsize $\Delta$}\rotatebox[origin=c]{-90}{$\longrightarrow$}
\\[6pt]
\begin{tabular}{|c|}
\hline
\makebox[32mm]{Supercond. class CI}
\\
\hline
$(3,1)$ \\
\hline
\end{tabular}
&
\xrightarrow{H}
&
\begin{tabular}{|c|}
\hline
\makebox[32mm]{Supercond. class C}
\\
\hline
$(2,0)$ \\
\hline
\end{tabular}
\end{array}
$$
\caption{Crossovers between spin-symmetric symmetry classes driven by the
magnetic field ($H$) and coupling to superconductors ($\Delta$).
The dimensions of the FF and BB sectors of the supersymmetric $\sigma$-model
for the average density of states are shown by $(n_F,n_B)$.}
\label{F:4classes}
\end{figure}

The purpose of this Letter is to theoretically study
crossovers between superconducting classes.

We will calculate the average density of states in a small 
diffusive metallic grain coupled to two superconducting 
terminals through tunnel barriers, see Fig.~\ref{F:setup}.
The terminals have the phase difference $\pi$
ensuring the absence of the minigap in the excitation spectrum \cite{pi}.
A magnetic field $H$ is applied to the system. The spin-rotation
symmetry is assumed to be intact.
We will be interested in the ergodic regime, $E\ll\ETh$,
where $\ETh=D/L^2$ is the Thouless energy,
$D$ is the diffusion constant, and $L$ is the grain size.

Under these conditions, the excitation spectrum in the grain
can be described in terms of the RMT in the crossover region
between the four symmetry classes shown in Fig.~\ref{F:4classes}.

\medskip

\textbf{Mapping to Efetov's $\sigma$-model.}
First attempts of field-theoretical description of hybrid NS systems
\cite{SKF98,Bundschuh99,Taras-JETPL,Taras-AdvPhys,OSF01}
inspired by the identification of superconducting symmetry classes \cite{AZ97}
have used the Bogolyubov--de Gennes (BdG) Hamiltonian as the starting point,
\be
  \hat{\cal H}_\text{BdG}
  =
  \begin{pmatrix}
    H & \Delta \\
    \Delta^* & -H^T
  \end{pmatrix}
  ,
\ee
where $H$ is the single-particle Hamiltonian,
and $\Delta(\br)$ is the pairing field.
The average quasiparticle DOS,
\be
  \corr{\rho(E,\br)} = - \Im \corr{\tr {\cal G}^\text{R}_E(\br)}/\pi ,
\ee
is expressed in terms of the retarded Green function of the BdG Hamiltonian,
\be
  {\cal G}^\text{R}_E = (E - \hat{\cal H}_\text{BdG} + i0)^{-1} ,
\ee
which is then represented as a functional integral
over an $8\times 8$ supermatrix field $Q$
acting in the direct product $\text{FB}\otimes\text{N}\otimes\text{PH}$
of the Fermi-Bose (FB), Nambu (N) and Particle-Hole (PH) spaces
(spin-symmetric case is considered).

In hybrid NS systems, Andreev reflection off the order parameter field $\Delta$
couples the states with opposite energies, $E$ and $-E$.
So, the Nambu-Gor'kov Green function ${\cal G}^\text{R}_E$
essentially involves a pair of the retarded and advanced
normal-metal Green functions, $G^\text{R}_E$ and $G^\text{A}_{-E}$.
In the absence of the superconducting pairing field, $\Delta(\br)$,
correlations between the latter
are conveniently described by Efetov's supersymmetric $\sigma$-model
\cite{Efetov1983,Efetov-book} of the orthogonal symmetry class,
with an $8\times 8$ superfield $Q$
acting in the direct product $\text{FB}\otimes\text{RA}\otimes\text{TR}$
of the Fermi-Bose (FB), Retarded-Advanced (RA) and Time-Reversal (TR)
spaces (again we assume no spin interactions).

Thus, in studying the proximity effect in the normal part of a hybrid system,
it is tempting to reformulate the field theory of Refs.\
\cite{SKF98,Bundschuh99,Taras-JETPL,Taras-AdvPhys,OSF01}
in the language of Efetov's supersymmetric $\sigma$-model.
Provided that the inverse proximity effect in the superconducting
regions
can be neglected (rigid boundary conditions),
Andreev scattering of normal electrons off
the superconducting terminal will be viewed as an effective boundary condition
at the NS interface mixing the R and A components of the field $Q$.
Such a description is close in spirit to the scattering
approach \cite{scattering}.

The average local DOS is given by the functional integral
over the normal-metal region \cite{OSF01}:
\be
\label{rho-def1}
  \corr{\rho(E,\br)}
  =
  \frac\nu4
  \Re
  \int
  \str (k \Lambda Q)
  e^{-S_D[Q]-S_\Gamma[Q]}
  DQ(\br)
  ,
\ee
where $S_D[Q]$ is the bulk action:
\be
\label{SD}
  S_D
  =
  \frac{\pi\nu}{8}
  \int d\br \,
  \str
  \bigl[
    D \left( \nabla Q + ie\mathbf{A}[\tau_3,Q] \right)^2
  + 4iE \Lambda Q
  \bigr] ,
\ee
and the action $S_\Gamma[Q] = S_{\Gamma_1}[Q] + S_{\Gamma_2}[Q]$
describes NS interfaces \cite{Efetov-book,bound-log}:
\be
\label{Sboundary}
  S_{\Gamma_a}
  =
  - \frac12
  \sum_i
  \str \ln [1+e^{-2\beta_i}Q_S^{(a)} Q^{(a)}] .
\ee
Here $\nu$ is the DOS per one spin projection at the Fermi level,
$\mathbf{A}$ is the vector potential,
$Q^{(a)}$ labels the $Q$ field at the boundary
with the $a$-th superconductor,
and the NS interface is specified by transmission coefficients
$T_i=1/\cosh^2\beta_i$, with $i$ labelling open channels.
The field $Q$ satisfies $Q^2=1$ and is subject to an additional
symmetry constraint
\be
\label{Q-self}
  Q=CQ^TC^T
  .
\ee

\begin{table}
\caption{Basic matrices in the two versions of the $\sigma$-model.}
\label{T:SN-E}
\begin{ruledtabular}
\begin{tabular}{ccc}
  & NS $\sigma$-model \cite{OSF01} & Efetov's $\sigma$-model \cite{Efetov-book} \\
\hline
space & $\text{FB}\otimes\text{N}\otimes\text{PH}$ & $\text{FB}\otimes\text{RA}\otimes\text{TR}$
\\
\hline
$\Lambda$ & $\sigma_z^\text{N} \sigma_z^\text{PH}$ & $\sigma_z^\text{RA}$
\\
$\tau_3$  & $\sigma_z^\text{N}$ & $\sigma_z^\text{TR}$
\\
$C$
& $  - \sigma_x^\text{N}
  \begin{pmatrix}
    i\sigma_y^\text{PH} & 0 \\
    0 & \sigma_x^\text{PH}
  \end{pmatrix}_\text{FB}
$
&
$ \sigma_z^\text{RA}
  \begin{pmatrix}
    i\sigma_y^\text{TR} & 0 \\
    0 & \sigma_x^\text{TR}
  \end{pmatrix}_\text{FB}
$
\\[12pt ]
\hline
$\hat\Sigma$ & $\sigma_x^\text{N}$
&
$ \begin{pmatrix}
    \sigma_1^\text{RA} \sigma_1^\text{TR} & 0 \\
    0 & \sigma_2^\text{RA} \sigma_2^\text{TR}
  \end{pmatrix}_\text{FB}
$
\end{tabular}
\end{ruledtabular}
\end{table}

In the NS $\sigma$-model for $\corr{{\cal G}_E^\text{R}}$ \cite{OSF01},
the matrices $\Lambda$, $\tau_3$ and $C$
are given by the first column of Table \ref{T:SN-E}.
An exact mapping to Efetov's $\sigma$-model
is realized by the similarity transformation
$Q\mapsto VQV^{-1}$ with the matrix
\be
  V
  =
  \begin{pmatrix}
  -1_\text{FB} & 0 & 0 & 0 \\
  0 & 0 & 0 & 1_\text{FB} \\
  0 & 1_\text{FB} & 0 & 0 \\
  0 & 0 & -k_\text{FB} & 0 \\
  \end{pmatrix}
  ,
\ee
where the inner (outer) grading corresponds to the PH (N) space,
and $k=\diag(1,-1)_\text{FB}$
(we follow notations of Ref.~\cite{Efetov-book}).
Conjugation by $V$ simultaneously transforms the matrices
$\Lambda$, $\tau_3$ and $C$ from NS representation to Efetov's representation
given by the last column of Table \ref{T:SN-E}.
This provides an exact mapping between the NS $\sigma$-model
for Green function of the BdG Hamiltonian, $\corr{{\cal G}^\text{R}_E}$,
to the standard Efetov's orthogonal $\sigma$-model
for the product $\corr{G^\text{R}_{E}G^\text{A}_{-E}}$.
On such a mapping both the
structure of the manifold and the $\sigma$-model action get reproduced.
We emphasize that this mapping takes place only in the normal part
of a hybrid NS system, where the pairing amplitude $\Delta=0$.

To complete the formulation of the model we have to specify the $Q$
matrix in the bulk of a superconductor, $Q_S$.
It has a familiar form parameterized
with the help of the spectral angle $\theta_S = \arctan(i\Delta/E)$ as
\be
\label{QS-S}
  Q_S
  =
  \Lambda \cos\theta_S
  +
  \hat\Sigma \sin\theta_S .
\ee
The most nontrivial ingredient of the mapping from the SN $\sigma$-model
to Efetov's $\sigma$-model is the form of the matrix $\hat\Sigma$.
In the initial NS representation \cite{OSF01} it is just the Pauli matrix
in the Nambu space: $\sigma_x^\text{N}$.
Conjugating by $V$ we get it in Efetov's representation:
\be
  \hat \Sigma
  =
  \begin{pmatrix}
    0 & \Sigma \\
    \Sigma^{-1} & 0
  \end{pmatrix}_\text{TR}
  ,
\qquad
  \Sigma
  =
  \begin{pmatrix}
    0 & k_\text{FB} \\
    1_\text{FB} & 0
  \end{pmatrix}_\text{RA}
  .
\ee
We see that \emph{superconducting boundary conditions violate
supersymmetry}: The matrix $\hat\Sigma$
acts as $\sigma_1^\text{RA} \sigma_1^\text{TR}$ in the FF block
 and as $\sigma_2^\text{RA} \sigma_2^\text{TR}$ in the BB block.
This is the reason why a nontrivial DOS can be obtained by
integration (\ref{rho-def1}) over the standard
orthogonal $\sigma$-model manifold.

\medskip

\textbf{Zero-dimensional limit.}
In the ergodic regime, $E\ll\ETh$, the functional integral (\ref{rho-def1})
is dominated by the zero mode, $Q(\br)=\const$.
We will be interested in the average global DOS
normalized by the inverse mean quasiparticle
level spacing $\mls = (2\nu V)^{-1}$:
\be
  \corr{\varrho(E)}
  =
  \mls \int \corr{\rho(E,\br)} \, d\br .
\ee
This quantity can be written
as an integral over a single $8\times8$ supermatrix $Q$:
\be
\label{sigma0D}
  \corr{\varrho(E)}
  =
  \frac{1}{8}
  \Re
  \int
  \str (k \Lambda Q)
  e^{-S[Q]}
  DQ
  ,
\ee
with the action consisting of three terms:
\be
\label{S}
  S[Q]
  =
  \frac{ix}{4}
  \str
  \Lambda Q
  -
  \frac{\cicmatrix}{4}
  \str
  (\tau_3 Q)^2
  +
  \frac{\ucmatrix}{8}
  \str (\hat\Sigma Q)^2
  .
\ee
Here $x=\pi E/\mls$,
and the symmetry breaking parameters $\cicmatrix$ and $\ucmatrix$
are given by
\be
\label{params}
  \cicmatrix
  =
  \pi\nu De^2 \int\mathbf{A}^2 \, d\br
  ,
\qquad
  \ucmatrix = \frac{G_A}{8}
  .
\ee
One can estimate $\cicmatrix\sim (\phi/\phi_0)^2/g$,
where $\phi$ is the flux through the grain, $\phi_0$
is the flux quantum, and $g\sim\ETh/\mls\gg1$ is the
dimensionless grain conductance.
The last term in the action (\ref{S}) is written in the tunneling
limit, $T_i\ll1$, and the parameter $\ucmatrix$ is expressed
through the dimensionless (in units of $e^2/h$)
Andreev conductance of the grain
\cite{Andreev-transmission}, $G_A = 2 \sum_i T_i^2$
(the factor 2 accounts for two NS interfaces assumed
to be identical, see Fig.~\ref{F:setup}).
A strong magnetic field randomizes electron phase and
the crossover from the unitary class to class C can be
obtained with just one superconducting terminal attached
to the grain (in that case $G_A=\sum_i T_i^2$).
The last term in Eq.~(\ref{S}) is written in the subgap limit,
$E\ll\Delta$, when $Q_S=\hat\Sigma$.

Equations (\ref{sigma0D}) and (\ref{S}) describe the average DOS
in the two-parametric crossover between the four symmetry
classes shown in Fig.~\ref{F:4classes}. Instead of studying
the complicated general behavior, we will restrict ourselves to two
one-parametric crossovers: class A -- class C ($\cicmatrix=\infty$,
$\ucmatrix$ arbitrary) and class CI -- class C ($\ucmatrix=\infty$,
$\cicmatrix$ arbitrary).

\begin{figure}[b]
\centerline{
\includegraphics{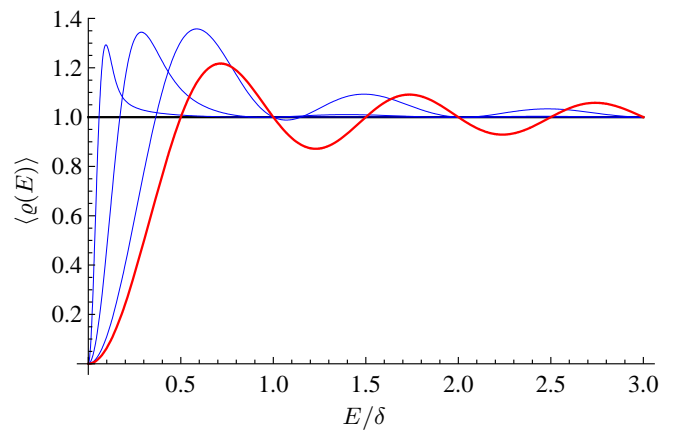}
}
\caption{The average DOS, $\corr{\varrho(E)}$, in the class A (unitary)
-- class C crossover. The curves correspond to different values
of the symmetry-breaking parameter:
$\ucmatrix=0$ (unperturbed DOS, $\corr{\varrho(E)}=1$), 0.01, 0.1, 1,
and $\infty$ (class C).}
\label{F:U-C}
\end{figure}

\medskip

\textbf{Class A -- class C crossover.}
In sufficiently strong magnetic fields ($\cicmatrix\gg1$),
cooperon degrees of freedom get frozen and the $Q$ matrix
becomes diagonal in the TR space:
\be
  Q
  =
  \begin{pmatrix}
    Q_u & 0 \\
    0 & k \Lambda Q_u^T \Lambda k
  \end{pmatrix}_\text{TR}
  ,
\ee
where the $4\times4$ matrix $Q_u\in\text{FB $\otimes$ RA}$
spans the manifold of the unitary Efetov's $\sigma$-model.
Therefore one can simply take the well-known Efetov's
parametrization of this manifold \cite{Efetov-book},
in which the FF and BB sectors are parametrized
by a radial variable ($\lambda_\text{F,B}$)
and an angular variable ($\varphi_\text{F,B}$),
with $-1\leq\lambda_\text{F}\leq1$ and $\lambda_\text{B}\geq1$.
It can be easily seen that the term $(\ucmatrix/8)\str(\hat\Sigma Q)^2$
in the action (\ref{S}) does not depend on $\lambda_\text{F}$.
Thus coupling to a superconductor suppresses only the BB degrees
of freedom, shrinking it to a point at $\ucmatrix\to\infty$
(see Fig.~\ref{F:4classes}).
Calculation of the integral (\ref{sigma0D})
is straightforward leading to the exact expression for the class A
-- class C crossover:
\be
\label{rho-U-C}
  \corr{\varrho(E)}
  =
  1
  -
  2 \ucmatrix \frac{\sin x}{x}
  \int_1^\infty d\lambda \, \lambda \, \cos{\lambda x} \, e^{-\ucmatrix(\lambda^2-1)}
  .
\ee
The function $\corr{\varrho(E)}$ is plotted in Fig.~\ref{F:U-C}
for several values of the symmetry breaking parameter $\ucmatrix$.

In the limit $\ucmatrix\to\infty$, one recovers the C-class result:
\be
\label{rho-C-class}
  \corr{\varrho_\text{C}(E)}
  =
  1
  -
  \frac{\sin 2x}{2x}
  .
\ee
In the limit of weak coupling to a superconductor, $\gamma\ll1$,
the uniform metallic average DOS is perturbed in a small vicinity
of the Fermi energy, at $E\lesssim\mls\sqrt\ucmatrix$,
but this perturbation is strong, completely suppressing the DOS
at $E=0$:
\be
  \corr{\varrho(E)}
  \approx
  f\left(\frac{x}{2\sqrt\gamma}\right),
\quad
  f(z) = 2z e^{-z^2} \int_0^z e^{t^2} dt .
\ee
Vanishing of $\corr{\varrho(0)}$ can be explained by repulsion
of energy levels $E$ and $-E$ which becomes effective
at very small energies $E\lesssim\mls\sqrt\ucmatrix$.
(Formally, noncommutativity of the limits $E\to0$ and $\ucmatrix\to0$
is a consequence of the noncompactness of the BB sector of the theory.)

Level statistics in classes A and C is known to be described
by free fermions: The joint probability density of energy levels
can be interpreted as the square of the ground-state wave function
for a system of noninteracting one-dimensional (1D) fermions.
Class A (unitary) corresponds to free fermions on a line \cite{Mehta},
while class C corresponds to free fermions with the Dirichlet boundary
condition at the origin \cite{AZ97}. Our result (\ref{rho-U-C})
indicates that in the crossover A--C energy levels cannot be considered
as noninteracting fermions in an appropriate single-particle potential.

\medskip
\textbf{Class CI -- class C crossover.}
Now we turn to the case of strong coupling to a superconductor,
$\ucmatrix\gg1$. The corresponding term in the action (\ref{S})
enforces $\str(\hat\Sigma Q)^2=0$. The latter condition
together with the relation $\{\Lambda,\hat\Sigma\}=0$
leads to the linear constraint
\be
\label{anticom-Q-Sigma}
  \{Q,\hat\Sigma\}=0.
\ee

The next step in constructing the parametrization suitable for
calculation in the crossover region is to study the commuting
generators $W$ of the CI-class manifold.
Writing $Q=\Lambda(1+W+\dots)$ with $\{W,\Lambda\}=0$,
and solving the linearized constraints (\ref{Q-self})
and (\ref{anticom-Q-Sigma}),
we find four commuting generators, three residing in
the FF sector, and one in the BB sector (see Fig.~\ref{F:4classes}):
\be
  W_\text{com}
  =
  \begin{pmatrix}
    W_\text{FF} & 0 \\
    0 & W_\text{BB}
  \end{pmatrix}
  =
  W_\text{com}^{(+)}+W_\text{com}^{(-)},
\ee
where
\be
  W_\text{FF}
  =
  \frac{i}2
  \begin{pmatrix}
    0&0&z&-c\\
    0&0&c&z^*\\
    z^*&c&0&0\\
    -c&z&0&0
  \end{pmatrix}_\text{RA}
  ,
\ee
\be
  W_\text{BB}
  =
  \frac{q}{2}
  \begin{pmatrix}
    0&0&0&1\\0&0&1&0\\0&1&0&0\\1&0&0&0
  \end{pmatrix}_\text{RA}
  ,
\ee
with complex $z$ and real $c$ and $q$.

As the $\cicmatrix$-dependent term in the action (\ref{S}) contains
the symmetry-breaking matrix $\tau_3$,
it is convenient to split $W_\text{com}$ into a part $W_\text{com}^{(+)}$
commuting with $\tau_3$ ($z$ and $z^*$ modes), and a part $W_\text{com}^{(-)}$
anticommuting with $\tau_3$ ($c$ and $q$ modes).
At finite $\cicmatrix$, the latter modes acquire a mass
proportional to $\cicmatrix$.
They will completely freeze out in the C-class limit ($\cicmatrix\to\infty$),
where only the modes $z=\theta e^{i\varphi}$ and $z^*=\theta e^{-i\varphi}$
will be unaffected, generating the sphere $S^2$ in the FF sector.

In constructing the global parametrization of the CI-class manifold
we will follow an approach of Ref.~\cite{AIE93} in order to maximally
simplify the symmetry breaking term
$S_\cicmatrix[Q] = -(\alpha/4)\str(\tau_3Q)^2$. We start with
parametrizing the commuting content of $Q$ as
$
  Q_\text{com}
  =
  U_\text{C}^{-1} U_\text{m}^{-1}
  \Lambda U_\text{m} U_\text{C}
$,
where
\be
\label{U-comm}
  U_\text{C} = \exp W^{(+)}_\text{com}
  ,
\qquad
  U_\text{m} = \exp W^{(-)}_\text{com}
  .
\ee
With such a choice, $S_\cicmatrix[Q_\text{com}]$ will explicitly
depend only on the massive modes $c$ and $q$.

Now we turn to the Grassmann content of the parametrization.
We search for Grassmann generators $\tilde W$ which obey
$[\tau_3,\tilde W]=0$ (to simplify the term $S_\alpha$)
and $[\Lambda,\tilde W]=0$ (to simplify the term $\str\Lambda Q$).
Employing also linearized Eqs.~(\ref{Q-self})
and (\ref{anticom-Q-Sigma}), we find that these anticommuting generator
parametrized by two Grassmann numbers:
\begin{gather}
  W_\text{Gr}^{(+)}[\xi,\rho]
  =
  \begin{pmatrix}
    u & 0 & 0 & 0 \\
    0 & v & 0 & 0 \\
    0 & 0 & v & 0 \\
    0 & 0 & 0 & -u
  \end{pmatrix}
  ,
\\
  u
  =
  \begin{pmatrix}
    0 & \xi \\
    -\rho & 0
  \end{pmatrix}_\text{FB}
  ,
\qquad
  v
  =
  \begin{pmatrix}
    0 & \rho \\
    \xi & 0
  \end{pmatrix}_\text{FB}
  .
\end{gather}

The desired parametrization of the $Q$ manifold in the spirit
of Ref.~\cite{AIE93} has the form
\be
\label{Q-param}
  Q
  =
  U_{\xi}^{-1} U_\text{C}^{-1} U_{\mu}^{-1} U_\text{m}^{-1}
  \Lambda U_\text{m} U_{\mu} U_\text{C} U_{\xi} ,
\ee
where the matrices $U_\text{m}$ and $U_\text{C}$ are defined
in Eq.~(\ref{U-comm}), and
\be
\label{U-anticomm}
  U_\xi = \exp W^{(+)}_\text{Gr}[\xi,\rho]
  ,
\qquad
  U_\mu = \exp W^{(+)}_\text{Gr}[\mu,\eta]
  .
\ee

After some algebra we obtain the Berezinian
of the parametrization (\ref{Q-param}):
\be
  J
  =
  \frac{\sin\theta}{2(1-\cos\theta)}
  \frac{\cos^2c}{(\sin c+i\sinh q)^2}
  \equiv
  J_\text{C} J_\text{m}
  .
\ee
Similar to Ref.~\cite{AIE93}, it splits into factors
depending either on massless or on massive coordinates.

The ingredients of the action (\ref{S}) take a simple form:
\begin{gather}
  \str(\tau_3 Q)^2 = 4(\cos 2c-\cosh 2q) ,
\nonumber 
\\
  \str\Lambda Q
  =
  - 4 \left[P + (1-\lambda) R \xi \rho\right] ,
\nonumber 
\end{gather}
where $\lambda=\cos\theta$, $P=\cosh q - \lambda \cos c$, $R=\cosh q-\cos c$,
while the preexponent in (\ref{sigma0D}) is the most involved:
\begin{multline}
  \str(k\Lambda Q)
  =
  4 \bigl[
    (\cosh q + \lambda \cos c) + 2P \, \eta\mu
\\{}
+ (1+\lambda) R \, \xi\rho + 2(1-\lambda) R \, \eta\mu\xi\rho
+ \dots
  \bigr],
\nonumber 
\end{multline}
where the omitted terms do not contribute to the average DOS.

Let us expand the integrand in Eq.~(\ref{sigma0D})
in Grassmann variables:
$$
F_{00}+F_{20}\,\xi\rho+F_{02}\,\mu\eta+F_{22}\,\xi\rho\mu\eta+\dots
$$
The parametrization (\ref{Q-param}) is singular
at $\theta=0$ ($U_\text{C}=1$) and
at $c=q=0$ ($U_\text{m}=1$).
Therefore the integral (\ref{sigma0D}) will contain
not only the regular contribution from the term $F_{22}$,
but also the contributions from the terms $F_{00}$, $F_{20}$ and $F_{02}$
which are finite due to the
Parisi--Sourlas--Efetov--Wegner theorem \cite{PSEW},
in complete analogy with the calculation of Ref.~\cite{AIE93}.
In particular, the terms with $F_{00}$ and $F_{20}$ reproduce
the C-class result (\ref{rho-C-class}),
while the other two terms are responsible for the crossover.
After some algebra, the general expression for the average DOS
in the class CI -- class C crossover takes the form

\begin{widetext}
\be
\label{rho-CI-C}
  \corr{\varrho(E)}
  =
  1
- \frac{\sin 2x}{2x}
+ \frac 1{2\pi}\int_{-\infty}^{\infty}dq\int_{-\pi/2}^{\pi/2}dc \,
  \cos c \, \frac{\sinh^2q-\sin^2c}{\sinh^2q+\sin^2c}
\sin(x\cos c)\sin(x\cosh q)\exp[\cicmatrix( \cos 2c-\cosh2q)].
\ee
\end{widetext}

\begin{figure}[b]
\centerline{
\includegraphics{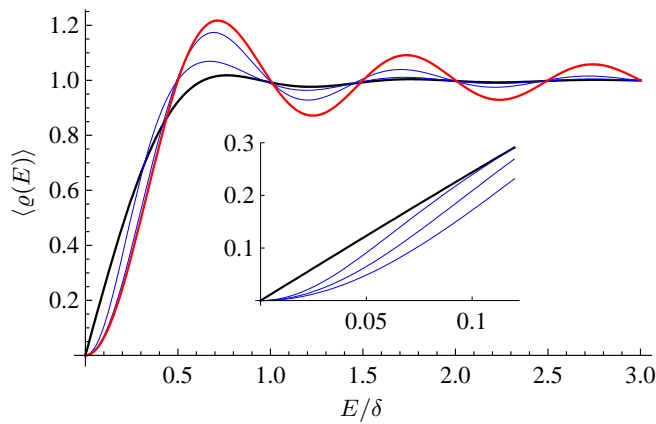}
}
\caption{The average DOS, $\corr{\varrho(E)}$, in the class CI
-- class C crossover. The curves correspond to different values
of the symmetry-breaking parameter:
$\cicmatrix=0$ (class CI, tiny oscillations),
0.2, 1, and $\infty$ (class C, large oscillations).
Inset: $\corr{\varrho(E)}$ for small deviations from class CI:
$\cicmatrix=0$, 0.01, 0.025, and 0.05.}
\label{F:CI-C}
\end{figure}

The DOS given by Eq.~(\ref{rho-CI-C}) is shown in Fig.~\ref{F:CI-C}
for several values of the parameter $\cicmatrix$.

In the limit $\cicmatrix=0$, Eq.~(\ref{rho-CI-C}) reproduces
the known result for class CI \cite{AZ97}:
\be
\label{rho-CI}
  \corr{\varrho_\text{CI}(E)}
  =
  \frac{\pi}{2}
  \left(
    x [J_0^2(x)+J_1^2(x)] - J_0(x) J_1(x)
  \right) .
\ee
A small magnetic field ($\cicmatrix\ll1$)
changes the linear CI-class behavior $\corr{\varrho(E)}\propto E$
to the quadratic C-class behavior $\corr{\varrho(E)}\propto E^2$
at $x\sim\sqrt\cicmatrix$:
\be
  \corr{\varrho(E)}
  \approx
  \frac{\pi}{4} x \erf \biggl( \frac{x}{\sqrt{8\cicmatrix}} \biggr) ,
\ee
see inset in Fig.~\ref{F:CI-C}.
A similar modification of the level repulsion exponent
takes place in the orthogonal--unitary crossover, 
Eq.~(\ref{pair-corr-O-U}).

In the limit of large $\cicmatrix$, one finds
\be
  \corr{\varrho(E)}
  \approx
  1 - \frac{\sin2x}{2x} \frac{4\cicmatrix}{\sqrt{16\cicmatrix^2+x^2}} ,
\ee
which reduces to the C-class result (\ref{rho-C-class})
at $\cicmatrix\to\infty$.

\medskip

\textbf{Conclusion.}
The purpose of this Letter was to study crossovers between
superconducting symmetry classes.
We have shown that the average DOS in the crossover
region between normal/superconducting symmetry classes with
spin-rotation symmetry (Fig.~\ref{F:4classes})
can be calculated using Efetov's supersymmetric
$\sigma$-model of orthogonal symmetry.
We have obtained exact expressions (\ref{rho-U-C})
and (\ref{rho-CI-C}) for the DOS
in the crossover regions between the classes A--C and CI--C.

We thank D. A. Ivanov and P. M. Ostrovsky for stimulating discussions.
This work was supported by the RFBR grant No.\ 10-02-01180,
the Dynasty Foundation,
and the Russian Federal Agency of Education (contract No.\ NK-529P).

\end{document}